# Automated Planning in Repeated Adversarial Games

**Enrique Munoz de Cote**
**Archie C. Chapman**
ECS, University of Southampton
{jemc,acc}@ecs.soton.ac.uk

**Adam M. Sykulski**
Imperial College London
adam.sykulski@imperial.ac.uk

**Nicholas R. Jennings**
ECS, University of Southampton
nrj@ecs.soton.ac.uk

## Abstract

Game theory's prescriptive power typically relies on full rationality and/or self–play interactions. In contrast, this work sets aside these fundamental premises and focuses instead on heterogeneous autonomous interactions between two or more agents. Specifically, we introduce a new and concise representation for repeated adversarial (constant–sum) games that highlight the necessary features that enable an automated planning agent to reason about how to score above the game's Nash equilibrium, when facing heterogeneous adversaries. To this end, we present TeamUP, a model–based RL algorithm designed for learning and planning such an abstraction. In essence, it is somewhat similar to R-MAX with a cleverly engineered reward shaping that treats exploration as an adversarial optimization problem. In practice, it attempts to find an ally with which to tacitly collude (in more than two–player games) and then collaborates on a joint plan of actions that can consistently score a high utility in adversarial repeated games.

We use the inaugural Lemonade Stand Game Tournament[1] to demonstrate the effectiveness of our approach, and find that TeamUP is the best performing agent, demoting the Tournament's actual winning strategy into second place. In our experimental analysis, we show hat our strategy successfully and consistently builds collaborations with many different heterogeneous (and sometimes very sophisticated) adversaries.

## 1 Introduction

Adversarial games[2] have been studied since the birth of game theory. They abstract many real scenarios such as chess, checkers, poker, and even the cold war. Repeated interactions, on the other hand, model situations where two or more agents interact multiple times, building long–run relationships. Their solutions are often represented as a sequence of actions (or plans of actions). Repeated adversarial games are, therefore, a natural framework to study sequential decision making in an adversarial setting. It is somewhat surprising, then, to find that artificial intelligence offers few insights on how to interact against heterogeneous agents in repeated adversarial games, particularly since the field is renowned for offering solutions under weaker assumptions than those imposed by game theory (e.g. regarding computation and memory power). Indeed, the problem is further exacerbated if the adversaries' behaviour is unknown *a priori* to the planning agent and the latter needs to interact with its adversaries to *learn* how they behave.

Against this background, we present a study on how an agent should plan a sequence of actions in a repeated adversarial game when the adversaries are *unknown*. Specifically, in these games, the capability of an agent to compute a good sequence of actions relies on its capacity to forecast the behaviour of its opponents, which may be a hard task against opponents that themselves *adapt*. The planner, therefore, needs to construct a plan while considering the *uncertain* effects of its actions on the decisions of its opponents.

The repeated nature of the interaction allows us to frame this as a planning adversarial problem, where the unknown behaviour of the opponents is learnt by repeatedly interacting with them. The learning task however, requires a large number of repeated interactions to be effective, which in many scenarios is unacceptable. To this end, we present a new and concise game abstraction that reduces the state space to a size that is tractable for an agent to learn with a small number of samples. Building on this, we present a reinforcement algorithm that learns on such a game abstraction. It is grounded on reward–shaping and model–based learning, techniques that have both been shown to decrease exploration complexity when compared with a straightforward implementation of Q–learning.

By so doing, our work deviates substantially from other work in multiagent learning against heteroge-neous opponents. First, instead of assuming that our planner is facing the worst opponent, as (Littman, 1994; Brafman & Tennenholtz, 2002) do, our objective is to design a planner whose resulting strategy can do better than its security (max–min) equilibrium strategy. Second, on the other extreme, our objective is not optimality against a known algorithm, as (Banerjee & Peng, 2005; Munoz de Cote & Jennings, 2010) consider. In that work, the authors derive planning solutions that are optimal only when facing some specific class of opponents. Our analysis, instead, does not put any type of constraints on the class of adversaries our planning agent might face. As a consequence, because an adversary may have any level of sophistication, in our setting it is not possible to produce optimal solutions.

---

[1] http://users.ecs.soton.ac.uk/acc/LSGT/home.html
[2] Adversarial games are also known as constant–sum games.

The paper is organized as follows: in Section 2 we give useful background and definitions. In Section 3 we define the planning problem more precisely and introduce our novel state abstraction. Following this, in Section 4 we propose an algorithm capable of learning such a state abstraction. In Section 5 we introduce the setting for our experiments, and in Section 6 we test the performance of our algorithm. Finally, in Section 7 we conclude.

## 2 Background and Definitions

Throughout this work we consider $n$ players that face each other and repeatedly play a normal form game.

A **normal form game (NFG)** is a $n$–player simultaneous–move game defined by the tuple $\Gamma = \langle \mathcal{N}, \{\mathcal{A}_i\}, \{r_i\}\rangle$, where $\mathcal{N}$ is the set of $n$ players, and for each player $i \in \mathcal{N}$, $\mathcal{A}_i$ is player $i$'s finite set of available actions, $\mathcal{A} = \times_{i \in \mathcal{N}} \mathcal{A}_i$ is the set of joint actions and $r_i : \mathcal{A} \to \mathbb{R}$ is the player $i$'s reward function. An agent's goal is to maximise its reward, and its **best response correspondence**, $BR_i(\mathbf{a}_{-i})$, is the set of $i$'s best strategies, given the actions $\mathbf{a}_{-i} \in \times_{j \in \mathcal{N} \setminus i} \mathcal{A}_j$ of the other players:

$$BR_i(\mathbf{a}_{-i}) = \{a_i \in \mathcal{A}_i \,:\, a_i = \operatorname*{argmax}_{a'_i \in \mathcal{A}_i} r_i(a'_i, \mathbf{a}_{-i})\}$$

In our context, stable points are characterised by the set of **pure Nash equilibria** (NE), which are those joint action profiles, $\mathbf{a}^*$, in which no individual agent has an incentive to change its action:

$$r_i(a_i^*, \mathbf{a}_{-i}^*) - r_i(a_i, \mathbf{a}_{-i}^*) \geq 0 \quad \forall\, a_i,\, \forall\, i \in \mathcal{N}. \quad (1)$$

That is, in a pure NE, $a_i^* \in BR_i(\mathbf{a}_{-i}^*) \,\forall\, i \in \mathcal{N}$. In many games, a player may find it beneficial to collaborate with some subset of the other players. Clearly, this occurs in coordination games; moreover, it can also arise in general–sum games (to choose from different equilibria) and even more surprisingly, in constant–sum games (we will explain how later in this section). To make this notion of collaboration precise, we now introduce a refinement to the best response correspondence above. First, note that if the size of an agent's best response set $|BR_i(\mathbf{a}_{-i})| > 1$, then there may exist an $a_i \in BR_i(\mathbf{a}_{-i})$ such that the payoff to an opponent $j$ is greater for this action than any other element of $i$'s best response. Call such a set of actions $i$'s *player $j$–considered* best response, $BR_{i \to j}(\mathbf{a}_{-i})$, given by:

$$r_j(BR_{i \to j}(\mathbf{a}_{-i}), \mathbf{a}_{-i}) \geq r_j(a'_i, \mathbf{a}_{-i}) \,\forall a'_i \in BR_i(\mathbf{a}_{-i}).$$

Given this, if, by $i$ playing a $j$–considered best response, $j$'s current action becomes an element of its best response set, then we call this a *reciprocal* best response. Specifically, if the following holds:

$$a_j \in BR_j(BR_{i \to j}(\mathbf{a}_{-i}), \mathbf{a}_{-ij}),$$

where $-ij = N \setminus \{i, j\}$, then we call $BR_{i \to j}(\mathbf{a}_{-i})$ an $i$–to–$j$ reciprocal best response, written $BR_{i \leftrightarrow j}(\mathbf{a}_{-ij})$. Furthermore, this refinement can be generalised to consider sets of players, $\chi \in \mathcal{N} \setminus i$, instead of single players $j$. Although such reciprocal best responses are not guaranteed to exist for every player and for each action profile in a game, such an action does exist at a Nash equilibrium of a game (by definition). Given this, the reciprocal best response concept will be used as a basic building block of the TeamUP algorithm.

In our setting, the agents play a **repeated game**, in which a NFG is repeatedly played. In this context, the NFG is called a *stage game*. At each time step, the agents simultaneously choose an action, giving a profile $\mathbf{a} \in \mathcal{A}$, which is announced to all players, and each player $i$ receives their reward $r_i(\mathbf{a})$ in the NFG. This stage game is repeated at each time step.

Repeated NFGs can have a *finite* or *infinite horizon*, depending on the number of stage games played. A player's objective in a repeated game is to maximize the sum of rewards received. Imagine that at time $t$, players have seen the **history** of play $(\mathbf{a}^1, \ldots, \mathbf{a}^t) \in (\mathcal{A})^t$. In this context, a **behavioural policy** is a function $\pi_i : (\mathcal{A})^t \to \mathcal{A}_i$ that maps histories to actions. However, computing an optimal policy at any time $t$ is an *adversarial optimization problem* of the objective function $\pi_i^* = \arg\max_{\pi_i} \sum_{k=t}^{T} r_i(\pi_i, \mathbf{a}_{-i}^k)$. It is called adversarial because the term $\mathbf{a}_{-i}^k$ (i.e. the adversaries' *optimized* strategy) is coupled with the term to be maximized, $\pi_i$ [3]. The capability of a player to compute an optimal strategy therefore relies on its capacity to *forecast* its counterpart's behaviour, which may be a hard task against opponents that adapt.

## 3 Planning in Adversarial Repeated Games

Computing a strategy of play in repeated NFGs is an adversarial optimization problem, as explained earlier. And the planner's optimization capabilities rely on how well it can predict the behaviour of its adversaries. Against adaptive opponents, however, predicting their behaviour means learning a mapping from histories to the opponent's response strategies. Nevertheless, adversarial games involving two or more players are endowed with some underlying structure that can help reasoning using a higher–level representation. Specifically, it is possible to balance the (constant) sum in the planner's favour by colluding with other player(s) in order to minimize the utility of the excluded adversaries. However, such collaboration is difficult to achieve without explicit communication, correlation or pre–game agreements, which we do not consider.

---
[3]A change in $\pi_i$ might cause a change in $\mathbf{a}_{-i}^k$.

We focus instead on tacit collusion, in which teams of collaborating agents are formed through repeated play. To implement this collusion, we adopt the approach of modelling the *high–level decision–making behaviour* model of our opponents, rather than focusing on their specific actions. As we show later, this abstraction allows us to accurately forecast the opponent reactions to our current actions, while at the same time it is simple enough for state transitions to be learnt in a very small number of iterations. Together, these two properties allow our planner to collaborate effectively, when such an opportunity exists.

### 3.1 State Abstraction

Our objective is to collaborate with other players by making them an offer they can't refuse, i.e. an action profile that consistently gives them a score above the game's Nash equilibrium. We do this by measuring the deviation of the planner's opponents from *ideal types*, which the planner can easily predict and collaborate with. In this work, we use the ideas on leading Q–learning algorithms (Littman & Stone, 2002) to generate our ideal types. We now formally describe these ideal types and then show how we generalise these by measuring opponents' deviations from ideal types, which creates instances of features. Finally, we will describe how the state space is constructed using a tuple of these features. It is this abstraction that will help us build an automatic planner in the spirit of Littman and Stone's work.

**Ideal types**

There are two obvious ways of initiating collaborations. First, a player could *lead* simply by sticking to its current strategy, and wait to see if any opponents follow. Second, a player could *follow* by changing strategy to one that is inside the BR set. Based on these two patterns of play, we define two ideal types of strategies that an agent could easily collaborate with, as follows:

- A perfect *Lead* strategy picks a starting strategy and does not move from it for the duration of the game.
- A perfect *Follow* strategy always selects actions that are a BR to the previous action selected by the opponent that is being followed.

**Features**

The planner classifies the opponents by their proximity to playing either a lead or follow strategy, based on their previous actions. An opponent classified as playing a lead strategy is usually slow moving, or stationary, and is hence very predictable. An opponent classified as playing a follow strategy tends to play actions that are within the best response set from the previous time step (or an appropriately discounted average of recent time steps). Given this, we now discuss these ideal types and how we measure an opponent's deviation from them, which form the basis for our state abstraction.

In order to classify its opponents, the planner maintains a measure of a lead index, $l_i$, and a follow index, $f_{ij}$, (where $j \in \mathcal{N} \setminus i$), to measure whether player $i$ is following player $j$. The indices are calculated from the sequence of past actions of each player $A_i = (a_i^1, \ldots, a_i^{t-1})$:

$$l_i = -\sum_{k=2}^{t-1} \frac{\gamma^{t-1-k}}{\Gamma} \Delta\left(a_i^k, a_i^{k-1}\right)^\rho, \qquad (2)$$

$$f_{ij} = -\sum_{k=2}^{t-1} \frac{\gamma^{t-1-k}}{\Gamma} \Delta\left(a_i^k, BR_i(a_j^{k-1})\right)^\rho, \qquad (3)$$

where $\Gamma = \sum_{k=2}^{t-1} \gamma^{t-1-k}$. The function $\Delta(a_i^k, BR_i(a_j^{k-1}))$ is a metric that captures the *distance* between the actions of player $i$ at time–step $k$ and the $BR_i(a_j)$ at $k-1$. These indices therefore quantify the closeness of each opponent to an ideal type by looking at the lag–one distance between respective action sequences. The parameter $\rho$ scales the distances between actions: $\rho < 1$ treats all behaviour that deviates from the ideal types relatively equally, while $\rho > 1$ places more value on behaviour that is close to ideal type behaviour. As such, with a $\rho > 1$, our planner can accommodate players that are noisy but select actions close to that of an ideal type. Notice that the indices are always negative — the greater the value of the index, the more this player follows the ideal type. An index value of 0, indicates an exact ideal type. The parameter $\gamma \in (0, 1]$ is the response rate (or discount factor), which exponentially weights past observations. Note that these metrics generalize the idea of "distance", for example, in games where actions are physical locations (as is the running example from Section 5), $\Delta(\cdot)$ can represent the Euclidean distance (if using the Euclidean norm) and in less structured games, for example where any two different actions are treated the same, $\Delta(\cdot)$ is a boolean 0 if the arguments are equal and 1 otherwise.

Now we describe how we use these metrics to generate features on a high–level state space so that a learning algorithm, whose objective is to find allies to collaborate with, can plan a sequence of high–level actions that can result in a higher than the NE expected utility in adversarial repeated games.

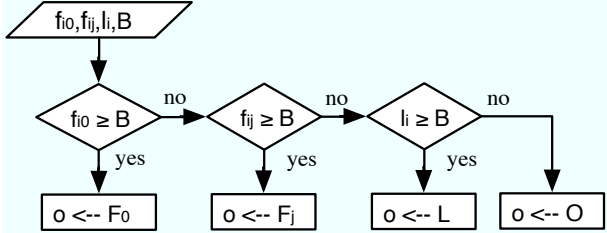

Figure 1: The decision flow of a state feature.

## 4 The TeamUP Algorithm

TeamUP is a model–based RL algorithm designed for learning and planning against unknown adversaries. In essence, it is reminiscent to R-MAX (Brafman & Tennenholtz, 2002) (in that it uses a model–based RL paradigm and implicit exploration) however it also incorporate a cleverly engineered reward shaping scheme that treats exploration as an adversarial optimization problem. In practice, TeamUP attempts to find allies with which to tacitly collaborate on a joint plan of actions that can consistently score a high utility in adversarial repeated games. Note that although we motivated this study on adversarial games, the basic properties of our algorithmic approach are general enough to be useful in the broader class of repeated general–sum games.

In more detail, using the metrics presented in the section, TeamUP creates *feature* instances, $o_i$, that take values from the set $\mathcal{O}_i = \{L, F_0, F_j, O\}$ (in the next subsection we explain what they mean in detail), and the flow chart in Figure 1 shows the decision process used to generate each instance. The input parameter $B$ is a threshold calculated as

$$B = f_{\min}\delta \quad (4)$$

where $f_{\min} = \min_{x,y \in A_i} -\sum_{k=2}^{t-1} \frac{\gamma^{t-1-k}}{\Gamma}\Delta(x,y)^\rho$ is a lower bound on the indices — note that $f_{\min}$ is an intrinsic parameter of the problem and indices $f_{ij}, l_i$ all take values in the range $[f_{\min}, 0]$. $B$ is solely "tweakable" by $0 \leq \delta \leq 1$ and in essence, this threshold modifies how tolerant to deviations from optimal types the planner is, with $\delta = 0$ being completely intolerant to deviations from optimal types.

### 4.1 States and Transitions

At a glance, the state representation of TeamUP is based on high–level observations that classify each opponent (and itself) as being either stationary ($L$), chasing another player ($F_x$), or unknown ($O$)[4]. Using this abstraction, the algorithm learns state transitions and expected rewards.

The planner's state is composed by a tuple of the form $s = (o_0, \{o_i\}_{i \in \mathcal{N}\setminus 0}) \in S = \times_{i \in \mathcal{N}}\mathcal{O}_i$ (which captures the high–level behaviour of all players in the game) and its action space is defined by the set $\{L, F_i, F_j\}$. From Eqs. (2,3), note how the discount factor $\gamma$ is what succinctly modifies how "high–level" strategies are treated. When $\gamma \to 0$, only immediate past actions are considered, and when $\gamma \to 1$ long sequences of past actions come into play. Also note that by structuring the game using this state abstraction, the planner is actually learning how to play a stochastic game, where state transitions are controlled by high–level strategies. TeamUP learns state–action transitions by counting experience triplets of the form $(s, a, s')$ (i.e. state, action and next state) and updating the transition function $T(s, a, s')$ when required[5]. Note that these transitions are (probably) non–stationary because the process is only partially controlled by TeamUP — the resulting next state also depends on the opponents' joint action. This contrasts with R-MAX ; It learns the transition function $T(s, \mathbf{a}, s')$, which are stationary transitions. Nevertheless, we aim to learn the (possibly non–stationary) transition function, because accurate transition predictions allow the planner to predict its opponents' responses and plan the best policy accordingly.

### 4.2 Social Reward Shaping Exploration

Explorative actions in a multiagent context deserve a different treatment to that of a single agent learning problem. This is because an agent's (typically random) exploratory actions, although *unintentional*, could be interpreted by its adversaries as deliberate strategic actions, so can contribute to the way its opponents react. This means that exploration in a MAL context is not only a heuristic that allows an agent to learn about its environment, but that it should also be cleverly designed to correctly signal its adversaries. Furthermore, explorative actions might be costly. This fact is exacerbated in finite horizon games where a player's chance to score high is lost forever.

Social reward shaping (Babes et al., 2008) are external rewards presented to a RL algorithm as an addition to the actual stage game rewards. Their purpose is to encourage desirable behavior during the learning process. In more detail, they are designed to try to "lock" non–stationary processes by acting as a leader in the early learning phase. They do so by reasoning about the changes in the players' strategies as a non–linear dynamic system, with probably many absorbing points (equilibria), and each with its basin of attraction. The resulting exploration heuristic that social shaping will induce (especially in the early learning phase) directly

---

[4] Note that we always use the index 0 for self referencing the planner.

[5] Note that the first element of the state tuple is the planner's last high–level action, which does not need to be measured.

'shapes' those basins of attraction, and therefore the probabilities that the system will converge to different equilibrium points.[6] Social shaping extends the well known potential–based shaping framework (Ng et al., 1999) to a multiagent context. Here, the system designer provides a real–valued function $\Phi : S \to \mathbb{R}$ and the potential function used to modify the reward for a transition from state $s$ to $s'$ is $F(s, s') = \gamma'\Phi(s') - \Phi(s)$, with the discount factor $\gamma'$.

When trying to extend potential based shaping to stochastic games (a generalization of Markov decision processes (MDPs) (Puterman, 1994) that allows multiple agents), the potential of a state (i.e. the state value) depends on the joint policy of all players, i.e.,

$$V^\pi(s) = r(\pi(s)) + \gamma' \sum_{s' \in S} T(s, \pi(s), s') V^\pi(s'). \quad (5)$$

Therefore, a high potential state for a certain joint policy $\pi$ might be a low potential state for some other joint policy $\pi'$ and there is no hope in defining an optimal joint policy for adversarial games. If this work was concerned with self–play analysis, or interested in strategies when facing fully rational opponents, a worst case solution for adversarial games would suffice (i.e. its max–min policy) and the state potentials would be well defined. However, we are interested in close to optimal policies against unknown adversaries, so state potentials need not be max–min based. Instead, what we define are potentials based on *ideal states*, that are played against ideal types of opponents.

To this end, we consider states, where at least one opponent is a perfect follower and is following the planner, as *optimal states*. For example, for a three player game, the set of optimal states is $\bar{S} = \{(L, F1, *), (L, *, F1), (F2, F1, *), (F3, *, F1)\}$, where $* \in \mathcal{O}$ is a wild–card feature. In the same way, the opposite applies to situations where the planner is left out from being followed as *worst states*, i.e. $\underline{s} \in \underline{S} = \{(*, F3, L), (*, L, F2), (*, F3, F2)\}$. Note how the optimal and worse states all express situations where tacit collusion between players exists. In adversarial games, the player(s) that are left out of the coalition will be the "sucker" players, achieving lower than max–min utilities. We'll build our planner around this fact, and therefore, its purpose will be twofold: make an offer the adversary cannot resist (find an ally), and exploit as effectively as possible the deficiencies of the sucker (left out) players in the planner's favour. Most importantly, note that states where collaborations exist (all states in the set $\{\bar{S}, \underline{S}\}$) are likely to be stable configurations, i.e. it is unlikely that there exists any profitable unilateral deviation[7]. For those states that we know have stable configurations, we can accurately define their state value. We build on this fact to work out our potential–based function. More specifically,

$$\Phi(\bar{s}) = V(\bar{s}) = \frac{R_{max}}{1 - \gamma'}$$

$$\Phi(\underline{s}) = V(\underline{s}) = \frac{R_{min}}{1 - \gamma'}$$

$$\Phi(\tilde{s}) = V(\tilde{s}) = \frac{R_{max} - \epsilon}{1 - \gamma'}$$

where $R_{max}$ and $R_{min}$ are the largest and smallest rewards respectively; $\tilde{s} \in S \setminus \{\bar{S}, \underline{S}\}$ are all other states and are $\epsilon$ lower than optimal, where $\epsilon$ can be chosen arbitrarily small to be optimistic in the face of uncertainty.

### 4.3 The Algorithm

TeamUP takes parameters $\gamma$, $\rho$, $\epsilon$, $\delta$ and $K$, and learns a model $M$ of the environment by experiencing tuples $\langle s, a, s', r\rangle$ and then computes an optimal policy with respect to this current model. Its execution can be divided in two phases:

**Initialisation:** Start with an initial estimate for the model parameters where all state–action pairs yield a reward based on assumptions from its specific shaping function (Section 4.2) and all states lead with probability 1 to the fictitious state $s_0$. Based on this current model, a call to VI($M$)[8] computes a new optimal discounted policy based on its current model $M$.

**For each stage game: (a) observe:** at state $s$, a new joint action $\mathbf{a} = (a, a_1, \ldots, a_n)$ is observed and the new features are computed using the decision flow (Figure 1), which constructs the next state $s'$. **(b) update:**

- $c(s, a, s') \leftarrow c(s, a, s') + 1$,
- $u(s, a) \leftarrow u(s, a) + r(\mathbf{a})$
- $R(s, a) \leftarrow \frac{u(s,a)}{\sum_{s'} c(s,a,s')}$,
- if $\sum_{s'} c(s, a, s') = K$, run VI($M$) and follow the new optimal discounted policy.

As can be seen, TeamUP is similar to R-MAX but differs in critical ways. A crucial difference is in the way TeamUP updates its model. It keeps counts of experienced tuples $\langle s, a, s'\rangle$, therefore, each time it runs VI, it recomputes the complete transition function $T(s, a, s'), \forall s, s' \in S, \forall a \in \mathcal{A}_i$, as opposed to R-MAX that only modifies the transition for the newly labeled 'known' tuple but not the rest. This is a crucial difference if opponents are non–stationary because state transitions depend on the adversaries' current strategies. The second difference is in the initialisation, where TeamUP uses social shaping to conduct its relaxation search, instead of using a fully optimistic relaxation search. As pointed before, a theoretical analysis of our algorithmic approach will not say much without fixing the type of opponents. We chose the Lemonade Stand Game Tournament (see next section) for driving our experimental analysis to test our initial premise — i.e. to design a planner that achieves high utilities against unknown adversaries. This tournament provided us the fairest neutral ground for comparison, beside providing us with very interesting and sometimes quite sophisticated adversary algorithms.

---

[6]A different approach for equilibrium selection is that of (Wicks & Greenwald, 2005), however that finds a unique stable equilibrium via perturbations, but not necessarily the one wanted.

[7]Under the assumption that left out players do not collude, the best these players can achieve is to optimize against their worst opponent (i.e. the planner and its ally). Any deviation from that max–min strategy results in a further advantage to the colluders.

[8]Where VI($M$) is a call to the standard value iteration algorithm (Puterman, 1994) on the current model $M$.

## 5 The Lemonade Stand Game

The Lemonade Stand game (LSG)[9] is played on an island, where, each day, three players choose a location for their lemonade stand, with the aim of being as far from their opponents as possible[10]. The game is played for 100 days — on each day the players choose their locations simultaneously and with no communication.

In the stage game, players choose from twelve possible locations, $A_i = \{1, \ldots, 12\}$. The total payoff sums to 24 and is to be distributed among the players given by the distance to the nearest player clockwise plus the distance to the nearest player anti–clockwise (i.e. customers are assumed to be uniformly distributed around the island). If two players are located in the same position, both receive 6 and the third receives 12. If all three are located in the same position, they all receive 8. Each player's objective is to maximise their aggregate utility over 100 rounds. As such, a player wishes to be located as far as possible from its two opponents in each stage game.

The stage game of the LSG has many pure NE: Figure 2 shows the NE locations for a third player, given players square and star are located as shown. For each configuration, the third player's best responses are anywhere in the larger segment between the star and square players, while the best responses that are consistent with a NE are those that are on or in–between the positions directly opposite the two players, as indicated by the arrows. This is clear in 2(a). In 2(b), where the opponents play opposite one another, the third player is indifferent between all positions, while in 2(c), where its opponents play on top of one another, the third player is indifferent between all positions except the location of its opponents. In particular, given the analysis above, it is unlikely that standard game theoretic analysis will suffice in building a good strategy. This motivates our introduction of $j$–considered and reciprocal best responses in Section 2. In particular, in any adversarial game played between three or more players, a coordinated team of players, $\chi \subset N$ can exploit the remainder $\chi'$. This means that in the LSG, two players can collaborate on a sequence of actions that maximize their utilities, at the expense of the third.

---
[9]The game was invented by Martin Zinkevich. It was designed to specifically test what should a good strategy be in repeated interactions between heterogeneous players, given that game theory's prescriptive power can only do so much in this domain.

[10]We refer the reader to (Sykulski et al., 2010) for a throughout description and analysis of the game.

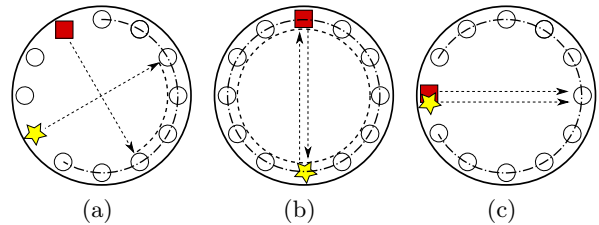

Figure 2: Best-responses for different opponent configurations: The dot–dashed segment indicates the third player's best–response actions, the dashed segment shows best–response actions consistent with a Nash equilibrium, and arrows point to the action opposite each opponent.

Specifically, collaboration between two players can be achieved using reciprocal best responses, and this forms the basis of TeamUP. In the LSG, a reciprocal best response is played by $i$ if it chooses to *follow* an opponent and play directly opposite it (these points are indicated by the arrows in Figure 2). In this situation, the utility of the third player is restricted to 6, which it receives in all of the 12 possible positions — hence all locations are NE, as shown in Figure 2(b). Thus, the two collaborating players share the remaining utility of 18. Similarly, player $i$ might choose to lead its opponents by *sticking* in one position and wait for another player to play a reciprocal best response to this action, with the same payoffs resulting. Note that a third strategy that selects locations randomly can easily be defeated this way — with the two collaborating players receiving an expected utility of 9 and the random strategy only 6. A pair of strategies that consistently play such reciprocal best responses can therefore frequently receive high payoffs.

## 6 Results

In this section we analyse and compare TeamUP to the other entries of the inaugural LSG Tournament, in order to demonstrate the performance of the planner's resulting strategy against other entrants of the original tournament. Testing TeamUP in this domain means our planner will face a pair of heterogeneous and unknown strategies in a more than two action adversarial game. The successful strategies will therefore be those that can consistently balance the constant sum in their favour.

## 6.1 The LSG Tournament

First, we re-ran the Tournament but included TeamUP (see Table 1). Note that we do not include results for Brown, which was placed 7th in the original tournament[11]. We repeat the structure of the original tournament, which is a round–robin format with each triplet combination of agents simulated for several repeats. The original Tournament concluded with $EA^2$ shown to be the winner and Pujara and RL3 awarded a statistical tie for second place. In the revised version, however, TeamUP is the overall winner, demoting $EA^2$ into second place. The parameter values that we used for TeamUP in all reported experiments are: $\gamma = 0.05, \delta = 0.3, \rho = 0.5, K = 15$ and these were chosen considering the game length. Also, note that for the LSG, $R_{max} = 12$ (for optimal states) and $R_{min} = 6$ (for worst states). We set $\epsilon = 4$ such that all other states have a potential of 8, which is a player's equal share of the total utility.

Table 1: LSG Tournament including TeamUP

| Rank | Strategy | Avg. Utility | S.E. |
|---|---|---|---|
| 1. | TeamUP | 8.5838 | ± 0.0098 |
| 2. | $EA^2$ (Southampton/Imperial) | 8.4635 | ± 0.0090 |
| 3. | RL3 (Rutgers) | 8.4267 | ± 0.0055 |
| 4. | Pujara (Yahoo! Research) | 8.4065 | ± 0.0082 |
| 5. | Waugh (Carnegie Mellon) | 8.1455 | ± 0.0106 |
| 6. | ACT–R (Carnegie Mellon) | 7.9356 | ± 0.0122 |
| 7. | Schapire (Princeton) | 7.5979 | ± 0.0120 |
| 8. | FrozenPontiac (Michigan) | 7.5382 | ± 0.0111 |
| 9. | Kuhlmann (UT Austin) | 6.9635 | ± 0.0076 |

There were several interesting strategies of note. The strategies placed 4th–9th, in our revised standings, all select actions similarly to an ideal type (lead or follow), and can therefore be simultaneously used as a collaborative partner (except for the random strategy used by Kuhlmann), or exploited as the sucker player. $EA^2$ and RL3, however, look to adapt their behaviour in order to guarantee forming a collaboration with a particular opponent. These two strategies share some underlying principles with TeamUP — in particular that actions should be selected according to the general behavioural characteristics of the opponents. It is of no surprise, therefore, that these strategies performed best in this Tournament. The key difference, and contribution of TeamUP, is that it learns reactions of the opponents to the planner's actions by using sequential planning to select collaborations that are predicted to yield the highest long–term utility dependent on the types of opponent faced. In contrast, $EA^2$ for example, is indifferent between types of collaboration and selects the statistically most likely partnership to succeed myopically, based on the recent behaviour of the opponents, without planning over the rest of the game. RL3 however, selects between potential collaborations based on their historical success, but similarly to $EA^2$, it does not do this by considering the fu-ture behaviour of both opponents — a feature which is inherent in the sequential planning performed by TeamUP.

---
[11]We apologise to the Brown team for this — their strategy is extremely complex and requires several weeks of computation to analyse, for which we had insufficient time.

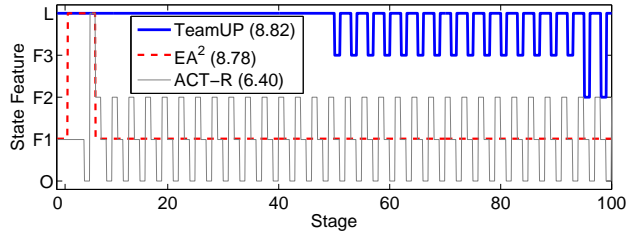

Figure 4: The classification of each player's state for an example game of 100 stages.

To gain yet more insight, we show some case examples which outline key reasons for TeamUP yielding the highest utility of all strategies. Specifically, in Figure 3 we show a breakdown of the most frequent visited states, aggregated over several repeats of the game, for various combinations of opponents. We classify together all stage game states that belong to the optimal state ($\bar{s}$), worst state ($\underline{s}$) or in all other state ($\tilde{s}$) sets. We also include the average utility received by each algorithm. From the figure, notice that 85–95% of the stage games reached belong to the optimal set. Although not noticeable in Figure 3, TeamUP's collaborative partners are $EA^2$ and RL3, which consistently follow TeamUP — this reinforces the claim that our strategy is consistent in finding an opponent to collaborate with and exploit the third player. Specifically, not only does TeamUP succeed in collaborating against different types of opponents, but the third player appears to be selecting actions near the opponent and far from our location, such that our utility is the highest of all three players.

To explain these results in more detail, in Figure 4 we show an example game run over its 100 stages and plot the high–level strategies played by the triplet at each stage. We choose ACT–R as this strategy cycles between strategies that include following one of the opponents, switching only if the utility received is below some acceptable level. In this game, TeamUP and $EA^2$ have successfully collaborated (as indicated by the average utilities), ACT–R therefore switches between states at every iteration, as its utility is consistently low. Crucially, TeamUP is identifying this change in behaviour from ACT–R. This fact follows from the fast learning time of TeamUP (due to our shaping rewards and efficient use of experience by the model–based RL) and that the response rate, $\gamma$, is set suitably low. Notice that occasionally TeamUP then chooses to follow ACT–R (TeamUP's feature $F3$), in an attempt by TeamUP to have both opponents play on the opposite side of the island (as $EA^2$ is already following TeamUP).

When taken together, the results in this section demonstrate the ability of TeamUP to not only form collaborations with various heterogeneous opponents, but to also often receive the higher proportion of the joint utility shared with the collaborative partner.

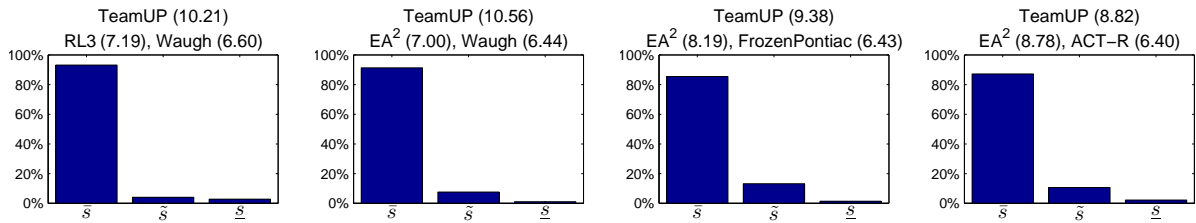

Figure 3: The frequency of visits to optimal states, $\bar{s}$, worst states, $\underline{s}$, and all other states, $\tilde{s}$, against various combinations of opponents, where we include the average utility to each player.

## 7  Conclusions and Future Work

This paper presents a novel way of thinking about multiagent heterogeneous interactions. Specifically, our analysis and results are sustained under the mathematical framework of repeated adversarial games. We propose a new way of analysing any repeated game played between the planner and one or more adversaries by abstracting important features about the interaction between players themselves. Specifically, this paper proposes that such features be defined in terms of "leaders" and "followers" and classifying the behaviour of our opponents under these terms. We introduced TeamUP to show how an automatic planning agent, by reasoning in a strategy space (as opposed to an action space), can generate policies that can score high utilities in adversarial games. Note that lead and follow strategies can be used in any repeated interaction game (not just adversarial) and hence our abstraction (and therefore TeamUP) should be general enough to work in these settings. However, our study is based on an adversarial setting given that this is the most challenging (due to the opposing players' goals).

Our experimental setting analysed and compared TeamUP to other entries in the inaugural LSG Tournament. Our findings, beside presenting TeamUP as the overall winner, revealed (surprisingly) clever policies that we did not have in mind. TeamUP showed not only that it is successful in building tacit collusion policies with an ally, but that at the same time, it encourages the sucker player to play close to a *planner*–considered BR (in the LSG this means to stay far way) — it is this extra component that significantly increased TeamUP's utility.

The study we present takes an important step forward in building good learning algorithms designed specifically to play against other unknown autonomous agents. Our proposed algorithm uses such an abstraction and, besides outperforming every other strategy experimentally, is (to our knowledge) the first automatic planner designed for any type of adversary. In the near future, we plan to work on the theoretical bound when facing different classes of adversaries. Also, it would be interesting investigating if our abstraction could be generalized with the idea of cluster-based representations found in (Ficici et al., 2008).